\documentstyle[12pt]{article}
\begin{document}
\title{Magnetic Field Effect in a Two-dimensional Array of
Short Josephson Junctions}
\author{G.Filatrella$^{a,b}$ and K. Wiesenfeld$^c$\\
\noindent
a)Department of Physics, University of Salerno,\\
I-84081 Baronissi (SA), Italy\\
b) Physikalisches Institut, Lehrstuhl Experimentalphysik II,\\
University of T\"ubingen, D-72076 T\"ubingen, Germany\\
c) School of Physics, Georgia Institute of Technology,\\
 Atlanta, Georgia, USA 30332\\}
\maketitle

\begin{center}
PACS number(s) 74.50+r, 74.60+Ge, 74.20+De, 05.40+j
\end{center}
\baselineskip=20pt
\begin{abstract}
We study analytically the effect of a constant magnetic field on the
dynamics of a two dimensional Josephson array.  The magnetic field
induces spatially dependent states and coupling between rows, even
in the absence of an external load.  Numerical simulations support
these conclusions.
\end{abstract}
\baselineskip=27pt

\section{Introduction}

Arrays of Josephson junctions
have attracted increasing attention in recent years.  One reason for
this
interest is the possibility of using Josephson arrays as millimeter-
wave
oscillators and
amplifiers. While single junctions might in principle be used for
such
purposes, in practice they deliver very little power when coupled to
an
external load {\bf \cite{likharev86}}.  Tilley proposed {\bf
\cite{tilley70}}  using 1D series arrays working coherently to match
typical
load impedances. Unfortunately
the mechanism of internal coupling of series arrays has proven to be
weak, thus requiring stringent conditions on the fabrication
tolerance {\bf \cite{wan89}}.

Two dimensional arrays have been investigated in the hope
that some internal mechanism might prove to be effective in
coherently phase locking the junctions.  Experimentally, encouraging
results have been reported on the emitted power
{\bf \cite{benz91}} and linewidth {\bf \cite{booi94}}. A heuristic
explanation for the success of 2D arrays is that the presence of
superconductive loops in the system provides a further coupling
mechanism among the junctions that is absent in 1D series arrays;
indeed evidence that fluxons (a $2 \pi$ wrap of the superconductor
phase trapped in a superconductive loop) do play a role in 2D arrays
has been reported
with the LTSEM technique {\bf \cite{lachenmann94}}. On the other
hand, theoretical analysis of {\em bare} 2D arrays ({\em i.e.} arrays
not coupled to an external load) in the absence of any magnetic field
shows that the uniform in-phase solution has similar neutral
stability features as bare 1D series arrays {\bf
\cite{hadley89,wiesenfeld94}}.  Indeed, recent simulations on one
class of disordered 2D arrays suggest that the external load is
responsible (in large part or entirely) for the coherent behavior
observed there {\bf \cite{kautz94}}.  These results point out the need
for a deeper fundamental understanding of
the dynamics of 2D arrays.  The inclusion of magnetic field effects
has perhaps been slowed by the fact that appropriate models of 2D
arrays in presence of magnetic field are in general rather
complicated {\bf \cite{phillips93,reinel94}}, so that a direct
theoretical attack on the problem is formidable.

The purpose of this paper is to build up some analytic insight
into the dynamics of fluxon states in two dimensional
arrays, especially the role played by
the magnetic field. Rather than study directly the general case of $M
\times N$ arrays (see Figure 1a), we focus on two simpler
configurations.  First, we consider the case of a single row of
plaquettes ({\it i.e.} a $2 \times N$ array, see Fig. 1b)
biased by a constant transverse current.
This is similar to the problem of a 1D parallel array studied
elsewhere{\bf \cite{cirillo93,doderer94}}, except that we
include junctions in the horizontal branches.  For
bias
currents not too close to the critical current, we construct fluxon
solutions whose spatial structure depends on the presence
of the magnetic field.  While this is somewhat enlightening,
the $2 \times N$ array is too simple to help us understand certain
important aspects of the two dimensional problem.  Thus,
we turn next to the case of a double row of
plaquettes ({\it i.e.} a $3 \times N$ array, see Fig. 1c.)
This is perhaps the
simplest arrangement which fully captures the essential features of
a
2D
array, insofar as it allows us to study the effect of weak
interactions
between
the rows.  We find that a fluxon state in the top row induces a fluxon
state in
the bottom row; moreover, the resulting dynamical state has
a definite phase relation between the fluxons in the two
rows,
in qualitative agreement with numerical simulations.  Remarkably,
{\em this relative phase becomes undetermined in the zero-field
limit}.  We conclude that the magnetic field breaks the symmetry
responsible
for the neutral stability found in bare 2D arrays {\bf
\cite{wiesenfeld94}}.
            \indent
\section{The model}

\indent
A two dimensional array of $N \times M$ short Josephson junctions
can be represented by the equivalent circuit depicted in Fig. 1a.
The most important effect we want to study is the
interaction between rows.  To this
end we consider a simplified model which includes only the self
inductances of each loop but ignores mutual inductances between
loops.  With this
simplification, following the usual analysis for superconductive
loops {\bf \cite{barone82}}, the equation of motion for this system
(in normalized units and for junctions of neligible capacitance, see
Fig.~1 for notation) are{\bf \cite{nakajima81,majhofer91}}

\begin{eqnarray}
\nonumber
\dot{V}_{l,j} &=& - \sin V_{l,j} +\gamma + \\
& &\frac{1}{\beta_l}
[V_{l+1,j} - 2V_{l,j} + V_{l-1,j}
+ H_{l,j} - H_{l,j+1} + H_{l-1,j+1} - H_{l-1,j}]\\
\nonumber
\dot{H}_{l,j} &=& - \sin H_{l,j} + \\
& &\frac{1}{\beta_l} [H_{l,j+1}
- 2H_{l,j} + H_{l,j-1} + V_{l,j} - V_{l+1,j} + V_{l+1,j-1} - V_{l,j-1}]\\
\end{eqnarray}
where $l=2, . . . , M-1$ and $j=2, . . . , N-1$.  Here, $V_{l,j}$ and
$H_{l,j}$ are the Josephson phase differences of the vertical and
horizontal junctions, respectively, $\gamma = I_B/I_0$ is the
normalized bias current,
$\beta_l = \frac{2\pi LI_0}{\Phi_0}$ is the usual SQUID parameter,
$\eta = \frac{\Phi^e}{LI_0}$ is the normalized external flux per
elementary cell ($\Phi^e$), $R$ and $I_0$ are the normal resistance
and the critical current of the junctions, respectively, $L$ is the
self
inductance of the superconducting loop, and the unit of time is
$\frac{\hbar}{2eRI_0}$.

The boundary conditions are ($N$ denotes the total number of
vertical
junctions
and $M$ the total number of horizontal junctions):
\begin{eqnarray}
\nonumber
\dot{V}_{1,j} &=& - \sin V_{l,1} +\gamma - \eta +
\\
& & \frac{1}{\beta_l} [V_{2,j} - V_{1,j} + H_{1,j} - H_{1,j+1}]
\;\;\;  \;\;\;  \;\;\;  j=1,...,M-1\\
\nonumber
\dot{V}_{N,j} &=& - \sin V_{N,j} +\gamma + \eta +\\
& & \frac{1}{\beta_l} [V_{N-1,j} - V_{N,j} - H_{N-1,j+1} - H_{N-1,j}]
\;\;\;  \;\;\;  \;\;\;  j=1,...,M-1\\
\nonumber
\dot{H}_{l,1} &=& - \sin H_{l,1} + \eta +\\
& & \frac{1}{\beta_l} [H_{l,2} - H_{l,1} + V_{l,1} - V_{l+1,1}]
\;\;\;  \;\;\;  \;\;\;  l=1,...,N-1\\
\nonumber
\dot{H}_{l,M} &=& - \sin H_{l,M} -\eta +\\
& & \frac{1}{\beta_l} [H_{l,M-1} - H_{l,M} - V_{l,M-1} + V_{l+1,M-1}]
\;\;\;  \;\;\;  \;\;\;  l=1,...,N-1
\end{eqnarray}

In this section and the next, we
consider the case of a single row of plaquettes which is current
biased in the transverse direction (Fig. 1b).  This system is similar
to the 1D parallel array of short Josephson junctions studied
elsewhere{\bf \cite{cirillo93,doderer94}}, but for the presence of
junctions in the horizontal branches. There are two horizontal
Josephson junctions for each vertical junction; however,
we can find solutions where the dynamics of the upper junction and
the lower junction are not independent, but rather satisfy

\begin{equation}
H_{l,1}(t) = -H_{l,2}(t) = H_l(t)
\end{equation}
where $H_{l,2}$ is the Josephson phase across the $l^{th}$ horizontal
junction in the upper branch and $H_{l,1}$ is the Josephson phase
difference
across the $l^{th}$ lower branch, an identity that allows us to
introduce the simplified notation $H_l$, as indicated.

To see that such dynamical states exist, add Eq.(5) to Eq.(6) with
$M=2$ to get
\begin{equation}
\dot{H}_{l,1} + \sin{H_{l,1}} = - \dot{H}_{l,2} - \sin{H_{l,2}}.
\end{equation}
\noindent
This is satisfied by $H_{l,1} = - H_{l,2}$ provided either (i) the
initial
conditions are the same for the two junctions or (ii) this is an
attracting state.  (Physically, we can arrange for identical initial
conditions if, before applying a driving current, we allow the system
to relax to the steady state $H_{l,1} = \dot{H}_{l,1} = 0$.)

We emphasize that our analysis relies on our neglecting the off
diagonal
terms of the inductance matrix {\bf \cite{phillips93}}: the
mutual inductances make the problem much more complicated.

Summarizing, our equations for the 1D row of plaquettes are:
\begin{eqnarray}
\dot{V_l} &=& - \sin{V_l} + \gamma + \frac{1}{\beta_l}[2(H_l-H_{l-
1})
+V_{l-1} -2V_{l}+V_{l+1}] ;\;\;\;  l=2,...,N-1\\
\dot{H_l} &=& - \sin{H_l} + \frac{1}{\beta_l}[V_l-V_{l+1}-2H_l] +
\eta;
\;\;\; l=1,...,N-1
\end{eqnarray}
where $V_l$ is the Josephson phase difference across the $l^{th}$
vertical junction.
The equations for the vertical
junctions at the left and right ends are:
\begin{eqnarray}
\dot{V_1} &=& - \sin{V_1} - \eta + \gamma + \frac{1}{\beta_l}[2H_1
 - V_1 +V_2]\\
\dot{V_N} &=& - \sin{V_N} + \eta + \gamma +
\frac{1}{\beta_l}[V_{N-1} -V_N-2H_{N-1}].
\end{eqnarray}
\noindent

\section{Approximate solution for the isolated row}

To find an approximate solution for the isolated row of plaquettes
we apply a scheme of successive approximations.  To first
approximation, we assume that the horizontal junctions are
completely inactive ($H_l \simeq 0$), as suggested by numerical
simulations of Eqs.(9-12) (see Fig. 2).  Under
this hypothesis the analysis is similar to that carried out for a
continuous (long) Josephson junction in Ref. {\bf
\cite{nagatsuma84}}.
For sufficiently large bias currents $\gamma >> 1 $, we can
approximate the solution as the sum of a linear term and a
small oscillating term:

\begin{equation}
V_l(t) = \theta_{l,0} + vt + X_l (t)
\end{equation}
where the $\theta_{l,0}$ and $v$ are constants to be determined
self-consistently.  Linearizing Eq.(9)
yields
\begin{eqnarray}
\nonumber
\dot{X_l}(t) = &-& v + X_l (t) + \sin{(\theta_{l,0}+vt)} +
\cos{(\theta_{l,0}+vt)}X_l(t)\\
&+& \frac{1}{\beta_l}\left[ \theta_{l-1,0} - 2\theta_{l,0} +
\theta_{l+1,0} +
X_{l-1}(t) - 2X_{l}(t) + X_{l+1}(t)\right].
\end{eqnarray}

After the balancing of the constants and assuming a
stationary wave profile $X_l(t) = A(t) e^{i(kl-\omega t)}$ we
obtain the following equation for the wave amplitude:

\begin{equation}
\dot{A}(t) = [i\omega + \cos{(\theta_{l,0} + vt)} +
\frac{2}{\beta_l}(\cos{k}
- 1)] A(t) + \sin{(\theta_{l,0}+vt)} e^{-i(kl-\omega t)}
\label{eq:amplitude}
\end{equation}
The solution of the associated homogeneous equation decays
exponentially with time, so except for transient behavior it is
sufficient to seek a particular solution of Eq.(\ref{eq:amplitude}).
We find

\begin{eqnarray}
V_l(t) &=& \theta_{l,0} +vt + A_v \sin{(\theta_{l,0} + vt)} + B_v
\cos{(\theta_{l,0} + vt)}\\
\theta_{l,0} &=& l\eta\beta_l
\end{eqnarray}
\noindent
where $A_v$ and $B_v$ are constants, and Eq.(17) is necessary to
satisfy the boundary conditions.
Notice that the solution does not contain the wave number $k$;
this parameter has canceled out.  This solution can be
viewed as a travelling wave in the sense that the equation
of motion in two adjacent cells is the same after a fixed time delay
$\Delta t =\beta_l\eta/v$.  The propagation velocity of the wave is
given by the physical distance between two cells divided by this
time.  It is important to note that no signal is in fact propagating
across the system: in fact the time delay between two junctions is
zero if the magnetic field is zero.  In other words, this is the
phase velocity of the wave rather than the group velocity.
A similar estimate for this velocity was derived in Ref.
{\bf \cite{reinel94}}.

The three parameters that appear in Eq. (16), namely $v$, $A_v$ and
$B_v$, are fixed by separately balancing the constant, sine, and
cosine terms in Eq. (9):
\begin{eqnarray}
B_v &=& 2(\gamma -v)\\
\frac{2 A_v}{\beta_l}(\cos{\eta \beta_l} -1) + v B_v &=& 1\\
v A_v - \frac{2 B_v}{\beta_l}(\cos{\eta \beta_l}-1) &=& 0
\end{eqnarray}
In the limit of very large $\gamma$ the solution is
\begin{equation}
v \simeq \gamma, \;\;\; A_v \simeq 0 \;\;\; B_v \simeq
\frac{1}{\gamma}.
\end{equation}
\noindent
In this limit the time delay is simply $\Delta t
=\beta_l\eta/\gamma$.  Although derived for bias currents $\gamma >> 1$,
this formal restriction is not required in practice.  For example, Fig. 3
compares the formula for the time delay and typical results from numerical
simulations with $\gamma = 3/2$.  The agreement is quite good.  This is
because the key approximation is that the oscillations are nearly sinusoidal,
which is valid as long as the bias current is not too close to the critical
current; of course the agreement improves with increasing $\gamma$.  From
Fig. 3 it is evident that the
formula systematically underestimates the actual value. This is
reasonable because the estimate is based on the approximation
$v=\gamma$, but this overestimates the velocity (a better
approximation is $v = (\gamma^2 -1)^{1/2}$).

The next step is to obtain an approximate solution for the dynamics
of the horizontal junctions. We proceed using the same
approximations as before, inserting into Eq. (10) the approximate
solution for the vertical junctions, Eqs. (16,17).
We write the solution as a constant plus an oscillating term:
\begin{equation}
H_l(t) = H_0 + Y_l(t)
\end{equation}
\noindent
and assume that the oscillating part is small ($Y_l << 1$).

Physically, the absence of a term which grows linearly in the time
(compare Eq.(13)) means that there is no d.c. voltage across the
horizontal
junctions ($<\dot{H_l}> = 0$). This is a reasonable assumption since
the
horizontal branches are unbiased; it is also what we find in the
numerical simulations.

Proceeding as for the vertical junctions we obtain for $H_i(t)$
a solution of the form
\begin{equation}
H_i(t) = H_0 + A_H \sin{(\theta_{l,0} + vt)} + B_H \cos{(\theta_{l,0}
+ vt)}
\end{equation}
where $H_0$, $A_H$ and $B_H$ are again to be determined using
harmonic
balance.
Balancing the constants fixes $H_0$ to be
\begin{equation}
\sin{H_0} = \frac{-2H_0}{\beta_l}.
\end{equation}
\noindent
The physical meaning of this constant is analogous to the classical
SQUID
phase shift induced by a magnetic field trapped in the loop {\bf
\cite{barone82}},
the
factor $2$ takes into account the fact that in this case there $4$
rather than $2$ junctions in the elementary loop.
In the limit of high bias current ($\gamma >> 1 $) and no trapped
magnetic field ($H_0 = 0$) we find
\begin{eqnarray}
A_H &=& \frac{1}{\gamma^2(\beta_l+2)^2}\left[ \frac{-
\sin{\eta\beta_l}}
{\beta_l + 2} +\frac{1}{\beta_l}(1-\cos{\eta\beta_l})\right]
-\frac{\eta\beta_l}{\gamma(\beta_l + 2)}\\
B_H &=& \frac{1}{\beta_l\gamma^3(\beta_l+2)}\left[ \frac{-
\sin{\eta\beta_l}}
{\beta_l + 2} +\frac{1}{\beta_l}(1-\cos{\eta\beta_l})\right]
\end{eqnarray}
\noindent
in agreement with our numerical simulations.  (For example, for
the same parameters as in figure 2, we find agreement to better than
20\%.)
Note that when the applied magnetic field vanishes ($\eta =0$),
Eqs.(25,26) give $A_H = B_H = 0$ so the
horizontal junctions are inactive. When the magnetic field is
present this is no longer true; nevertheless, the amplitude of the
oscillations for the horizontal junctions are much smaller than
those of the vertical junctions.

At this stage one could carry the analysis further, inserting the
solution (23) back into Eq. (9), and repeating the harmonic balance
procedure to get a more accurate expression for the phases
$V_l(t)$ and the
velocity $v$, then iterating the scheme for the horizontal junctions,
and so
on.  However, for our purposes the estimates Eqs. (18-20) and Eqs.
(24-26)
are adequate.  Let us summarize the main results of this section:
\begin{itemize}
\item[1)]For $\gamma >> 1$ the solution of the 1D array can be
approximated analytically by retaining only the first Fourier
component.  This approximation is a common one which has been
used in
previous
studies of a single junction.
\item[2)]In this limit there is a clear difference between horizontal
and vertical junctions, which results from the anisotropic bias
current:  the
vertical junction phases increase without limit (on average linearly in
time), while the
horizontal junction phases oscillate about a fixed value.
\item[3)]The magnetic field is responsible for the spatial non-
uniformity of
the dynamics: if the applied magnetic field is zero then the vertical
junctions
oscillate synchronously [$\theta_{l,0} = 0$ for all $l$, compare Eq.
(10)]
and the
horizontal junctions are inactive.
\end{itemize}
 \section{Coupling between two rows}

\indent
We now extend the analysis to the case of two rows of plaquettes.
Our main
interest is to study the interactions between rows.  To do this, we
proceed as
follows:  the solution for the first row is assumed to coincide with
the
solution for the isolated row, and with this imposed we solve the
equation
for the second row.  In other words we seek the solution of one
row driven by the unperturbed solution of the other row.  Although
this
scheme is "undemocratic", it has the virtue that the dynamical
equations are
tractable using the same approximations as in the last section.
The
equation for the second row reads:
\begin{eqnarray}
\nonumber
\dot{V_{l,2}} = -\sin{V_{l,2}} &+& \frac{1}{\beta_l}\big[ V_{l+1,2}
-2V_{l+1,2} +V_{l-1,2} \\
\nonumber
&+& (A_H - A_H\cos{(\eta\beta_l)} +
B_H\sin{(\eta\beta_l)})\sin{(V_{l,0}+vt)} \\
&+& (B_H - A_H\sin{(\eta\beta_l)}
- B_H\cos{(\eta\beta_l)})\cos{(V_{l,0}+vt)}\big] + \gamma
\end{eqnarray}
whose asymptotic solution, in the same sense discussed for the
single
row, is:
\begin{equation}
V_{l,2} = \theta_{l,0} + \delta + vt +
\overline{A}_v\sin{(\theta_{l,0}
+\delta+vt)} +
\overline{B}_v\cos{(\theta_{l,0}+\delta+vt)}.
\end{equation}
Again, the parameters $\delta$,$\overline{A}_v$, and
$\overline{B}_v$
are determined by a set of algebraic equations which result from
harmonic balance, namely
\begin{eqnarray}
\overline{B}_v &=& 2(\gamma+v)\\
\nonumber
\sin{\delta}\left\{ -\overline{A}_vv + \frac{1}{\beta_l}
\left[2\overline{B}_v(\cos{(\eta\beta_l)}-1)\right]\right\} &=&
\cos{\delta}\left\{\overline{B}_vv-1 +
\frac{2\overline{A}_v}
{\beta_l}
\left[\cos(\eta\beta_l)-1\right]\right\} +\\
& &
\frac{1}{\beta_l}\left[ A_H(1-\cos(\eta\beta_l)) +
B_H\sin(\eta\beta_l)\right]\\
\nonumber
\sin{\delta}\left\{- \overline{B}_vv + 1 -
\frac{2\overline{A}_v}{\beta_l}
\left[\cos(\eta\beta_l)-1\right]\right\} &=&
\cos{\delta}\left\{ -\overline{A}_vv + \frac{1}{\beta_l}
\left[2\overline{B}_vv(\cos{(\eta \beta_l)}-1)\right]\right\} +\\
& &
\frac{1}{\beta_l}\left[ B_H(1-\cos(\eta\beta_l)) +
A_H\sin(\eta\beta_l)\right].
\end{eqnarray}
The most striking feature of these equations is that if either $\eta
=0$
(no magnetic field) or $\beta_l \rightarrow 0$ (uncoupled limit)
then
$\delta$ is undetermined, {\em i.e.} the phase shift between the two
rows is arbitrary.  In other words the observed value of $\delta$
can be anything, and depends on the initial conditions.  On the
contrary, even a tiny magnetic field leads to
the selection of a specific value of $\delta$.  To
check that this conclusion is not an artifact of our approximation
scheme we have performed numerical simulations of the full
dynamical equations for a two-row array, and we have indeed
found (see Fig. 4) that in the presence of a magnetic field the final
asymptotic value of the phase shift $\delta$ is zero, regardless of
the initial conditions and also regardless of the values of the
parameters $\gamma$ and $\beta_l$.  Thus, while the
simulations qualitatively support our analysis, quantitatively
they do not: Eqs. (29-31) predict an asymptotic value
of $\delta$ that is parameter dependent and not, in general, equal to
zero (see Fig. 4).  We suspect that this disagreement in the value of
$\delta$ is an artifact of our separation of the system into a "slave"
row and a "master" row, and that an analysis which treats the two
rows on an equal footing would lead to a more accurate value of the
phase shift.

\section{Discussion and Conclusion}

Our analysis has shown that it is possible to
induce a spatially dependent solution in 2D arrays
by the application of a magnetic field. As a consequence
the magnetic field produces a coupling between rows, even in the
absence of an
external load.  In the limit of
zero magnetic field the phase shift between rows (for the simplest
case of two rows) becomes arbitrary, which signals that the
dynamics is only neutrally
stable.

This finding may have practical importance for the
application of Josephson junction arrays as local oscillators.
In the absence of an external load, it is known that (in the context of
lump
circuit equations) the inphase state of 2D arrays is neutrally
stable\cite{hadley89,wiesenfeld94}.  Neutrally stable
dynamical
states (other than the inphase state) also occur in a
variety of 1D series arrays both with and without external
loads \cite
{hadley87,hadley88,nichols92,strogatz93,watanabe94,wiesenfeld94,braiman94}.
One drawback to neutrally
stable dynamics is their intrinsic sensitivity to noise. As a result,
it is desirable to modify these arrays in a way which will stabilize
the dynamics ({\em i.e.} make the target dynamical state a {\em
bona fide} attractor).  One way to do this is to couple the array to an
appropriate external load, but this can have the disadvantage of
limiting the frequency range over which the array can
operate\cite{wiesenfeld94}.  An alternative possibility suggested by
the
present work is to induce coupling via the application of a magnetic
field.  Of
course, this also makes the dynamics spatially non-uniform, which
may itself be a drawback for applications.

We reiterate that our conclusions are based on a number of assumptions:  we
have i) included only self-inductances; ii) assumed that the junction
parameters are identical; iii) neglected the effects of any external
(parasitic)
load; and iv) ignored higher harmonics in the junction oscillations.  The
virtue of these assumptions is that, within the context of the idealized model,
we have achieved some level of analytic understanding of a notoriously
complex nonlinear system.  One direction for future theoretical work is to
extend our analysis to include these other effects.  Of these, the presence of
a
parasitic load and higher harmonics could be handled straightforwardly
within the same framework.  In contrast, the presence of mutual inductance
and disorder ({\it i.e.} variations among the junction parameters) is more
difficult;
nevertheless, we can make an educated guess as to how they might modify
the dynamics.

Consider first the role of mutual inductances.  Physically, mutual inductance
provides a mechanism for coupling non-neighboring junctions in a similar
manner as does the self-
inductance of a single loop.  This is reflected in the governing dynamical
equations:
self-inductance introduces a next-nearest-neighbor coupling; mutual
inductance will introduce further couplings whose strength, however,
diminishes with distance.  Thus, we expect the inclusion of mutual
inductance to increase the net coupling strength between junctions, thereby
providing
addtional interactions between rows which are responsible for breaking the
inherent neutral stability.  However, we expect nothing fundamentally new
in the observed dynamical behavior.

Turning next to the effects of having non-identical junctions in the array, we
can learn from some recent theoretical studies on
bare\cite{octavio92,landsberg95} two-
dimensional arrays in the absence of a magnetic field. These show that
disorder spontaneously induces shunt currents (transverse to the direction
of the imposed bias current) which tend to compensate for the mismatch
between junctions within a row; however, the inter-row dynamics remains
neutrally stable.  These findings are consistent with those of Kautz on
disordered 2D arrays with an external load\cite{kautz94}, who also found
that disorder
apparently plays an unimportant dynamical role in coupling rows.
Consequently, we expect that small amounts of disorder will not greatly
change the dynamical behavior we have described.

Of course, pulling together these various effects within a single theoretical
framework is a challenging task. On the other hand, within the context of the
idealized model studied here, we have achieved some level of analytic
understanding of these complex nonlinear dynamical systems.

\section{Acknowledgement}

\indent
We wish to thank Y. Braiman, T. Doderer, R.P. Huebener,
S.G. Lachenmann, and B. Larsen and especially S. Benz and R.L.
Kautz for useful comments and discussions.
The work was partially supported a grant from the U.S. Office of
Naval Research under contract number N00014-J-91-1257.  GF
wishes to thank Georgia Tech for their hospitality during the
preparation of
this work and the EU for financial support through the  Human
Capital and Mobility program (Contract No $ERBCHRXCT920068$).
\medskip
\par

\section*{Figure Captions}
\begin{itemize}
\item[Fig.1] Schematic circuit model for a) a two-dimensional array;
b) a single row of plaquettes; c) two rows.
\item[Fig. 2]$\dot{V}_i(t)$ and $\dot{H}_i(t)$ for a 1D row.
Parameters of the
simulations are: $N = 10$, $\beta_l = 4$, $\eta = \pi/4$, $\gamma =
1.5$.
\item[Fig. 3]Time delay of the voltage peak between two adjacent
cells compared
with theoretical estimate.
Parameters of the simulations are:
$N = 10$, $\beta_l = 1$, $\gamma = 1.5$.
\item[Fig.4]Time evolution of two vertical junctions in the same
column for
a)$\eta=0$ and b)$\eta = \pi/4$.
Parameters of the simulations are:
$M=3$, $N = 10$, $\beta_l = 1$, $\gamma = 1.5$.
\end{itemize}

\end{document}